\documentstyle[epsfig]{aipproc}

\begin{document}
\title{Precision of inflationary predictions and recent CMB anisotropy data}

\author{
Dominik J. Schwarz$^*$\thanks{
Financially supported by the Austrian Academy of Sciences.} 
\thanks{{\tt dschwarz@hep.itp.tuwien.ac.at}}, 
J\'er\^ome Martin$^{\dagger}$\thanks{{\tt jmartin@iap.fr}}, and 
Alain Riazuelo$^{\dagger}$\thanks{{\tt alain.riazuelo@obspm.fr}}}
\address{
$^*$Institut f\"ur Theoretische Physik, Technische Universit\"at Wien,
1040 Wien, Austria\\
$^{\dagger}$Institut d'Astrophysique de Paris, 98 boulevard Arago, 
75014 Paris, France}
\maketitle

\begin{abstract}
Inflationary predictions of the cosmic microwave background anisotropy
are often based on the slow-roll approximation. We study 
the precision of these predictions and compare them with the recent 
data from BOOMERanG and MAXIMA-1.
\end{abstract}

\section*{Introduction}

High quality measurements of anisotropies in the cosmic microwave background 
(CMB) probe the cosmic fluctuations generated during an inflationary  
epoch in the very early Universe \cite{fluc}. Recently, BOOMERanG 
\cite{BoomerangD} and MAXIMA \cite{MaximaD} teams announced the clear detection
of a first acoustic peak at an angular scale $\simeq 1^{\circ }$, which 
confirms the most important prediction of inflation: the Universe seems to 
be spatially flat \cite{analysis}. Another generic prediction of inflation 
is that the primordial spectra of density perturbations and gravitational
waves are {\it almost} scale-invariant. More CMB precision measurements will be 
available soon.

We argue \cite{MS2} that CMB predictions on the basis of the simplest 
inflationary model, slow-roll inflation \cite{slowroll}, are not as precise 
as could be believed from the accuracy of the power spectra \cite{Grivell}. 
We compare the predictions from the slow-roll approximation 
\cite{SL} with the exact solutions from the model of power-law inflation 
\cite{AW}. We find unacceptable large errors in the predictions of multipole
moments. The reason is as follows: The primordial spectrum is best approximated 
at some pivot scale $k_*$. A small error in the spectral index gives 
rise to a large error at wavenumbers that differ significantly from $k_*$, 
due to a large lever arm. A natural choice for the pivot scale is
the present Hubble scale, but leads to extremely large errors for high 
multipole moments. A shift of the pivot scale to the scale of  
the first acoustic peak decreases these errors dramatically (see Figure 
\ref{fig1}).  
 
In \cite{MRS} we compare the improved (optimal pivot scale) slow-roll 
predictions with recent CMB data (see Figure 2). Most data analysis so far 
\cite{analysis} was based on a power-law shape of the primordial spectra. 
This shape is {\em not} predicted by the slow-roll approximation, only the 
first two terms in a Taylor expansion with respect to the wavenumber coincide. 

\section*{Precision of slow-roll predictions}

Slow-roll inflation is very simple and is an attractor for many inflationary
models. Inflation driven by a single field
$\varphi$, may be characterized at a given moment of time $t_*$ by the 
parameters $\epsilon \equiv - [\dot{H}/H^2]_*$, 
$\delta \equiv - [\ddot{\varphi}/(H\dot{\varphi})]_*$, 
$\xi \equiv [(\dot{\epsilon} - \dot{\delta})/H]_*$, \dots, where $H$ is the 
Hubble rate. The condition for inflation is $\epsilon < 1$, whereas 
slow-roll inflation is characterized by $\epsilon \ll 1, |\delta| \ll 1,
\xi = {\cal O}(\epsilon^2, \epsilon\delta, \delta^2)$, and negligible higher
derivatives. Based on these approximations the power spectrum of the Bardeen 
potential $\Phi$ and of the amplitude of gravitational waves $h$ reads 
\cite{SL,MS2}
\begin{eqnarray}
\label{specsrd}
k^3P_{\Phi} &=& \frac{9 H_*^2 l_{\rm Pl}^2}{25 \pi \epsilon} 
\biggl[1-2\epsilon -
2(2\epsilon -\delta )\biggl(C+\ln \frac{k}{k_*}\biggr)\biggr], \\
\label{specsrgw}
k^3P_h &=& \frac{16  H_*^2 l_{\rm Pl}^2}{\pi}
\biggl[1-2\epsilon \biggl(C+1+\ln \frac{k}{k_*}\biggr)\biggr],
\end{eqnarray}
where $C\equiv \gamma _{\rm E}+\ln 2-2\simeq -0.7296$, 
$\gamma _{\rm E}\simeq 0.5772$ being the Euler constant. The pivot scale 
is defined as $k_* = (aH)_*$. Fixing $k_*$ corresponds to a choice of 
the time $t_*$ during inflation.
The spectral indices can be obtained from $n_{\rm S}-1\equiv {\rm d}\ln
(k^3P_{\Phi})/{\rm d}\ln k = - 4 \epsilon + 2 \delta$ and 
$n_{\rm T}\equiv {\rm d}\ln (k^3P_h)/{\rm d}\ln k = - 2 \epsilon$. 
We call this the next-to-leading order of the slow-roll approximation
(at the leading order strictly scale-invariant spectra are predicted).

\begin{figure}[t!] 
\setlength{\unitlength}{0.55\linewidth}
\begin{picture}(2,0.6)
\put(0.5,0.05){\makebox(0,0){$\ell$}}
\put(0.22,0.535){\makebox(0,0){$n_{\rm S} = 0.6$}}
\put(0.405,0.46){\makebox(0,0){$n_{\rm S} = 0.7$}}
\put(0.575,0.36){\makebox(0,0){$n_{\rm S} = 0.8$}}
\put(0.63,0.28){\makebox(0,0){$n_{\rm S} = 0.9$}}
\put(0.74,0.185){\makebox(0,0){$n_{\rm S} = 0.95$}}
\put(0.71,0.53){\makebox(0,0){Error in \% }}
\put(0.45,0.33){\makebox(0,0){\epsfig{figure=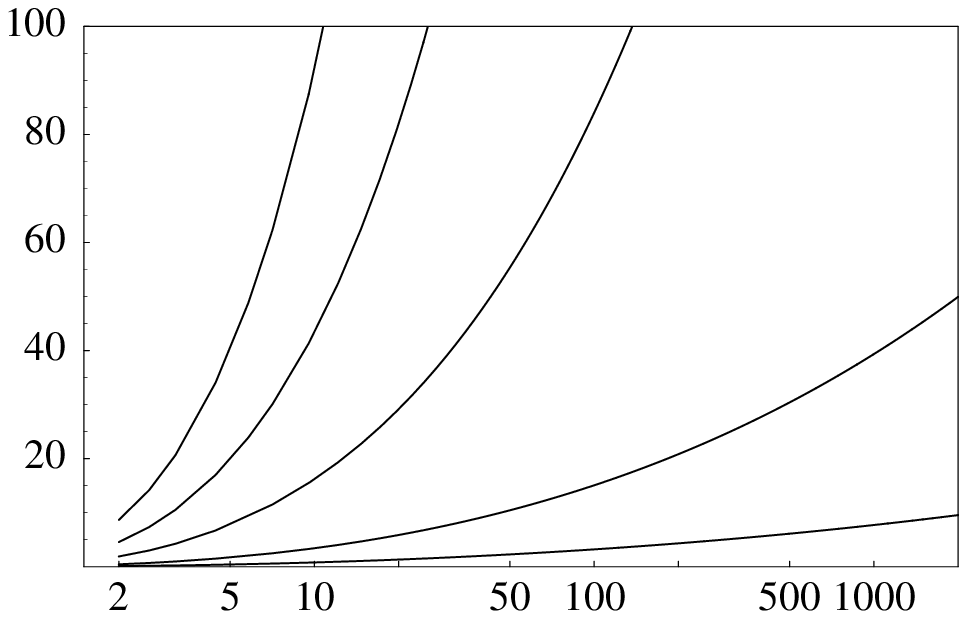,width=0.46\linewidth}}}
\put(1.4,0.05){\makebox(0,0){$\ell$}}
\put(1.13,0.40){\makebox(0,0){$n_{\rm S} = 0.6$}}
\put(1.08,0.325){\makebox(0,0){$0.7$}}
\put(1.08,0.255){\makebox(0,0){$0.8$}}
\put(1.08,0.19){\makebox(0,0){$0.9$}}
\thinlines
\put(1.06,0.14){\line(3,1){0.3}}
\put(1.47,0.245){\makebox(0,0){$n_{\rm S} = 0.95$}}
\put(1.61,0.53){\makebox(0,0){Error in \% }}
\put(1.35,0.33)
{\makebox(0,0){\epsfig{figure=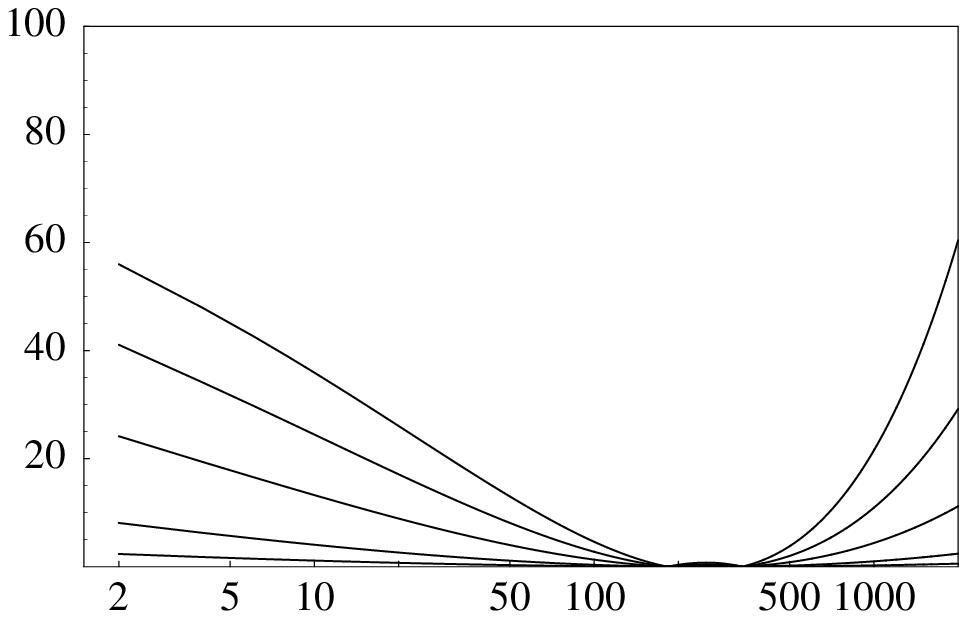,width=0.46\linewidth}}}
\end{picture}
\caption{The error of the scalar multipole moments 
$C_\ell$ from the slow-roll approximation w.r.t.\ the exact solution 
of power-law inflation. On the left side we show the errors with a pivot 
scale $k_*$ near the quadrupole, on the right side the pivot scale has been 
set to the scale of the first acoustic peak.}
\label{fig1}
\end{figure}

On the other hand, the power spectra may be calculated exactly for power-law
inflation, which is characterized by a power-law behavior of the scale factor,
i.e., $a \propto t^p$. For power-law inflation we have $\epsilon = \delta$
and $\xi=0$ during inflation. We use $\epsilon$ to parametrize the 
spectra, i.e. $p = 1/\epsilon$. The corresponding power spectra then read 
\cite{AW,MS}
\begin{equation}
\label{specpl}
k^3P_{\Phi}=\frac{9 H_*^2 l_{\rm Pl}^2}{25\pi\epsilon} f(\epsilon) 
\biggl(\frac{k}{k_*}\biggr)^{-\frac{2\epsilon}{1-\epsilon}},\
k^3P_h= \frac{16 H_*^2 l_{\rm Pl}^2}{\pi} f(\epsilon)
\biggl(\frac{k}{k_*}\biggr)^{-\frac{2\epsilon}{1-\epsilon}},
\end{equation}
where $f(\epsilon) = [2/(1-\epsilon)]^{2/(1-\epsilon)} 
\Gamma[1/(1-\epsilon) + 1/2]^2/\pi$, with $f(0) = 1$.
For power-law inflation the spectral indices read: $n_{\rm S} = 1 + n_{\rm T} 
= (1-3\epsilon)/(1 - \epsilon)$. In the limit $\epsilon \ll 1$ the power 
spectra (\ref{specpl}) go to (\ref{specsrd}) with $\epsilon = \delta$ and 
to (\ref{specsrgw}), respectively.  

We can now calculate the multipole moments $C_\ell$ for the power-law and 
slow-roll spectra for $\epsilon = \delta$. We define the error from the 
slow-roll approximation as
\begin{equation}
\label{deferr}
e_{C_\ell}\equiv 
\biggl\vert \frac{C_\ell^{\rm sr}-C_\ell^{\rm pl}}{C_\ell^{\rm pl}}\biggr\vert 
\times 100 \%\ .
\end{equation}
For similar spectra the error (\ref{deferr}) depends only weakly on the 
transfer function. This allows us to neglect the evolution of the transfer
function for this purpose and to obtain an analytic result, which is plotted
in Figure \ref{fig1}. The values of $n_{\rm S}$ refer to the exact 
power-law solution. In the left figure $k_* = k_0 \equiv (aH)_0$ 
gives the smallest error
for the quadrupole and unacceptably large errors at high multipoles.
In the right figure the pivot scale has been chosen to minimize the 
error around the first acoustic peak, $\ell \sim 200$. The corresponding 
condition is $D_{\ell_{\rm opt}}=\ln(k_*r_{\rm lss})$, where $r_{\rm lss}$ 
is the comoving distance to the last scattering surface and $D_{\ell} \equiv 
1-\ln 2+\Psi (\ell)+(\ell +1/2)/[\ell (\ell +1)]$ with $\Psi (x)\equiv 
{\rm d}\ln \Gamma (x)/{\rm d}x$. For $\ell_{\rm opt} \gg 1$ this gives 
$k_* \simeq (e\ell_{\rm opt})/(2 r_{\rm lss})$, where $r_{\rm lss} \simeq 
3.3/(aH)_0$ for $\Lambda$CDM. 

\section*{Slow-roll inflation and CMB anisotropy data}

\begin{figure}[t!] 
\centerline{\epsfig{file=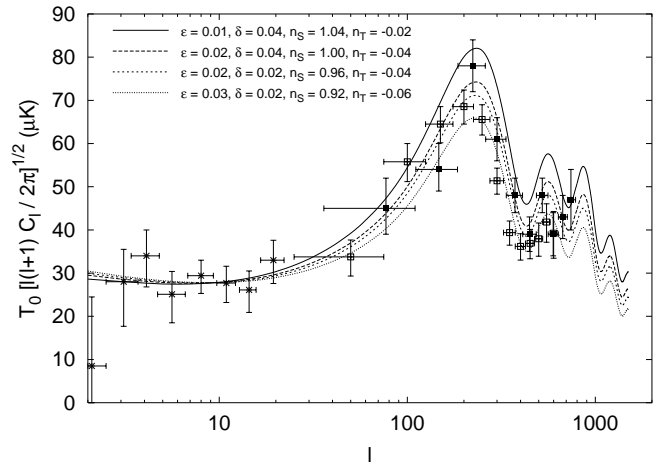,angle=270,width=0.61\linewidth}}
\vspace{10pt}
\caption{CMB band powers for slow-roll inflation in the $\Lambda$CDM scenario
for different values of the slow-roll parameters together with the
data points of the COBE/DMR (crosses), BOOMERanG (open boxes) and
MAXIMA-1 (filled boxes) experiments.}
\label{fig2}
\end{figure}

Let us now compare \cite{MRS} the predictions of slow-roll inflation with 
recent data from BOOMERanG \cite{BoomerangD} and MAXIMA-1 \cite{MaximaD}, 
supplemented with the COBE/DMR dataset \cite{COBE}. 
Instead of fitting ten cosmological parameters we fix the values of
otherwise measured parameters and assume that slow-roll inflation is the 
correct theory. In Figure \ref{fig2} we present the sum of scalar and 
tensor CMB band power for a $\Lambda$CDM model with 
$\Omega = 1, \Omega_\Lambda = 0.7, \Omega_{\rm b} h^2 = 0.019,$ and $h = 0.6$.
The Boltzmann code used here was developed by one of us (A.R.).
We see without a $\chi^2$ analysis that qualitatively different inflationary
models 
are consistent with the observations: Both models with $\epsilon = 0.02$ 
give reasonable fits, one of these models has a flat scalar spectrum 
(with $n_{\rm S} \neq n_{\rm T} + 1$), the other one has a negative 
tilt (with $n_{\rm S} = n_{\rm T} + 1$). Both models have an important 
contribution of gravitational waves ($\sim 20 \%$). 

We emphasize that the generic slow-roll predictions (\ref{specsrd}) and 
(\ref{specsrgw}) do not have a power-law shape. This fact induces large 
differences to multipole moments that are predicted under the assumption 
that the power-law shape (\ref{specpl}) is the generic inflationary 
prediction. Besides using the correct primordial spectrum a clever choice 
of the pivot scale can hide unavoidable uncertainties of the multipole moments 
in the cosmic variance on one side and in the instrumental noise 
on the other side of the spectrum.

\end{document}